# Universal sensitivity of speckle intensity correlations to wavefront change in light diffusers

KyungDuk Kim,[1] HyeonSeung Yu,[1] KyeoReh Lee,[1] and YongKeun Park[1*]

[1]*Department of Physics, Korea Advanced Institute of Science and Technology, Daejeon 34141, Republic of Korea*

Here, we present a concept based on the realization that a complex medium can be used as a simple interferometer. Changes in the wavefront of an incident coherent beam can be retrieved by analyzing changes in speckle patterns when the beam passes through a light diffuser. We demonstrate that the spatial intensity correlations of the speckle patterns are independent of the light diffusers, and are solely determined by the phase changes of an incident beam. With numerical simulations using the random matrix theory, and an experimental pressure-driven wavefront-deforming setup using a microfluidic channel, we theoretically and experimentally confirm the universal sensitivity of speckle intensity correlations, which is attributed to the conservation of optical field correlation despite multiple light scattering. This work demonstrates that a complex media is a simple interferometer, and presents opportunities to replace complicated reference-beam-assisted interferometers with a simple and compact scattering layer in various applications in metrology, analytical chemistry, and biomedicine.

## I. INTRODUCTION

Speckle is a granular pattern that appears when highly coherent light is scattered by a random structure [1]. In contrast to efforts to suppress speckle to improve imaging quality [2-5], there have also been approaches to exploit speckle patterns as a means of gathering physical information about a target. This field of research, known as speckle metrology, enables the measurement of various physical quantities such as vibration [6], roughness [7], and velocity [8]. At the same time, speckle has also been used to measure the optical characteristics of light, including wavelengths [9-11] and images [12,13].

One of the advantages of speckle metrology is its utility in phase-measurement interferometry [14]. Using speckle to detect changes in a wavefront enables measurement of the angular tilt [15] or surface deformation [16] of an object illuminated with coherent light. It has also been applied in the design of various kinds of optical remote sensors such as those used for gauging displacement [17], strain [18], and temperature [19]. More directly, the phase data of a beam can be retrieved from speckle using holographic methods [20,21] or iterative algorithms [22].

Despite the high resolution gained from the broad range of spatial frequencies in speckle, the implementation of phase-sensing methods requires both an interferometric setup with precise alignment and careful analysis of the diffraction patterns [23]. Recently we proposed that those requirements could be eliminated with a simple approach: scrambling a diffracted beam from an object by inserting a diffusive layer and then observing the decorrelation in the scattered fields, which can be used to quantify the wavefront deformation [24]. Without requiring the use of an interferometer, multiple scattering inside a diffuser conveys the phase change of the incident beam to the output speckle. The principle enables a phase-sensitive sensor with a simple geometry, which only requires adding a diffusive layer. Similar concepts of deformation sensing through scattering media have also been implemented in temperature [25] and position sensing [26]. Although multiple scattering in turbid media plays a critical role in these approaches, the effects and properties of scattering media have not been thoroughly investigated yet.

In this letter, we demonstrate the universal response of the diffusive scattering media to the phase changes of an incident light field. We numerically and experimentally study how the spatial intensity correlation coefficient of a speckle field after a diffusor is solely determined by the wavefront of a beam impinging on the diffuser, and the existence of a universal relationship regardless of the inserted diffuser.

The basic principles of the phase sensitivity of speckle patterns and its universal behaviors over diffusive layers are illustrated in Fig. 1. The main aim of speckle metrology is to measure phase deformations. However, these phase distortions are not observed in the intensity patterns, thus resulting in indistinguishable output patterns [Fig. 1(a)]. However, when the beam is scattered by a diffusive layer, its distorted wavefront forms speckle via interference. Because the scattered field completely depends on the phase of incident light, the speckle pattern is altered due to the difference between the initial and deformed wavefronts. In other words, the scattering layer itself serves as a simple type of interferometry: the phase change of the incident field is revealed through the change in the intensity profile of the output speckle.

Even though the phase distortion can be efficiently converted into highly uncorrelated intensity patterns via a diffuser, it is clear that the exact intensity patterns depend on the properties of the scattering layer. This fact can be problematic in applications for phase sensing because the behavior of the speckle pattern cannot be predicted without *a priori* information, or calibrated information about the scattering layer is required. Somewhat intriguingly, a universal relationship exists between the phase change of an incident light and the correlated spatial intensity after a scattering layer, which is independent of the properties of the scattering media [Fig. 1(b)]. A detailed confirmation will be presented in the following sections.

---

[*] yk.park@kaist.ac.kr

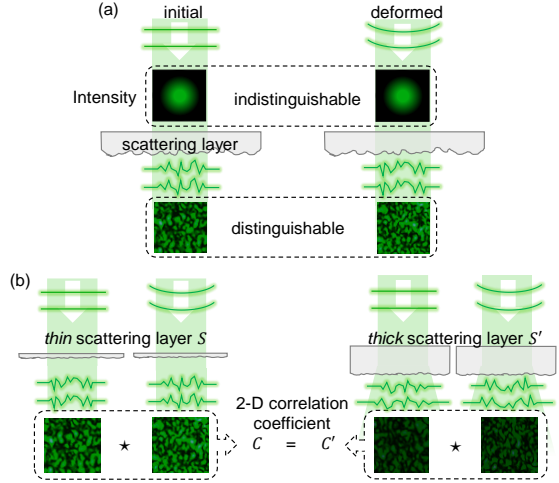

FIG. 1. (a) The effects of a scattering layer, which converts an indistinguishable phase difference between incident fields to the distinguishable changes in speckle patterns. (b) Speckle pattern changes for scattering layers of different thickness. When the change is expressed as a correlation coefficient, the values are the same regardless of the properties of the scattering layers, such as thickness. ★ symbol represents a cross-correlation of intensity maps.

## II.  MATHEMATICAL FORMULATION

We first mathematically formulate the intensity cross-correlation coefficient $g_{\mathbf{r}}^{(2)}$ between two output speckle patterns $I_a$ and $I_b$, $g_{\mathbf{r}}^{(2)} = I_a \star I_b$. Let $\mathbf{r}'$ and $\mathbf{r}$ be the coordinates at the input and output facets of a scattering layer. Now, assume that two monochromatic field $x_1(\mathbf{r}')$ and $x_2(\mathbf{r}')$ impinge onto the scattering layer, and corresponding scattered fields are $y_1(\mathbf{r})$ and $y_2(\mathbf{r})$, respectively.

Using a Green's function $G(\mathbf{r}, \mathbf{r}')$ to describe the scattering layer related to the incident and output fields, the exact form of the output scattered field is given as,

$$y_{1,2}(\mathbf{r}) = \int_{\mathbf{r}'} d\mathbf{r}' G(\mathbf{r}, \mathbf{r}') x_{1,2}(\mathbf{r}'). \qquad (1)$$

Because an image sensor measures the scattered intensity image rather than field the intensity correlation, $g_{\mathbf{r}}^{(2)}$ is calculated as,

$$g_{\mathbf{r}}^{(2)} = \frac{\langle y_2^* y_2 y_1^* y_1 \rangle_{\mathbf{r}}}{\langle y_2^* y_2 \rangle_{\mathbf{r}} \langle y_1^* y_1 \rangle_{\mathbf{r}}}, \qquad (2)$$

where $\langle \cdot \rangle_{\mathbf{r}}$ represents an ensemble over $\mathbf{r}$ space.

In order to progress beyond this point, an important assumption is required: all the scattered field exhibit complex Gaussian distribution over $\mathbf{r}$ space. Then, according to the Reed's moment theorem [27], Eq. (3) can be expressed as

$$g_{\mathbf{r}}^{(2)} = 1 + \frac{|\langle y_2^* y_1 \rangle_{\mathbf{r}}|^2}{\langle y_2^* y_2 \rangle_{\mathbf{r}} \langle y_1^* y_1 \rangle_{\mathbf{r}}}, \qquad (3)$$

which is the well-known Siegert relation applied in spatial domain [28]. By substituting Eq. (1) into $\langle y_2^* y_1 \rangle_{\mathbf{r}}$ term we can directly connect the scattered field correlation with incident field correlation via autocorrelation of Green's function,

$$\langle y_2^* y_1 \rangle_{\mathbf{r}} = \iint_{\mathbf{r}', \mathbf{r}''} d\mathbf{r}' d\mathbf{r}'' \langle G^*(\mathbf{r}, \mathbf{r}'') G(\mathbf{r}, \mathbf{r}') \rangle_{\mathbf{r}} x_2^*(\mathbf{r}'') x_1(\mathbf{r}'). \qquad (4)$$

Since $\langle G^*(\mathbf{r},\mathbf{r}'')G(\mathbf{r},\mathbf{r}')\rangle_\mathbf{r}$ can be regarded as $\langle |G|^2 \rangle_\mathbf{r} \delta(\mathbf{r}''-\mathbf{r}')$ for full rank transmission geometry with sufficiently large $\mathbf{r}$ space, Eq. (4) can be simplified as

$$\langle y_2^* y_1 \rangle_\mathbf{r} = A_{\mathbf{r}'} \langle |G|^2 \rangle_\mathbf{r} \langle x_2^*(\mathbf{r}') x_1(\mathbf{r}') \rangle_{\mathbf{r}'}. \quad (5)$$

where $\langle |G|^2 \rangle_\mathbf{r}$ is a constant ranged in [0,1], and $A_{\mathbf{r}'}$ represents $\int_{\mathbf{r}'} d\mathbf{r}'$. Please note similar calculation also holds for $\langle y_1^* y_1 \rangle_\mathbf{r}$ and $\langle y_2^* y_2 \rangle_\mathbf{r}$ terms. Substituting Eq. (5) into Eq. (3), for all second order moments of scattered fields we get

$$g_\mathbf{r}^{(2)} = 1 + \left| g_{\mathbf{r}'}^{(1)} \right|^2, \quad (6)$$

where $g_{\mathbf{r}'}^{(1)}$ represents the normalized first-order correlation function of input fields over $\mathbf{r}'$ space. Please note that Eq. (6) shows $g_\mathbf{r}^{(2)}$ only depends on the change in the incident fields, not the properties of the scattering layer. In other words, *any* diffusive scattering layer that satisfies the Reed's theorem, the speckle correlation directly represents the field correlation of incident fields, which verifies the field sensing ability of speckle metrology.

### III. NUMERICAL SIMULATION AND RESULTS

In order to identify when the scattering layer would satisfy the Reed's theorem, we employed a numerical simulation which emulates the transmission matrices (TMs) of scattering media. The sets of TMs of scattering media was obtained with an approach used in disordered metallic systems [29] and more recently, in light transport in turbid media [30-32].

The scattering strength of a medium was controlled by changing the optical thickness $L/l_s$, or the ratio of the thickness of a layer $L$ to its scattering mean free path $l_s$. In our numerical simulations, a TM relates 1024 input modes of the incident field to 1024 output modes of a scattered field in the diffusive regime.

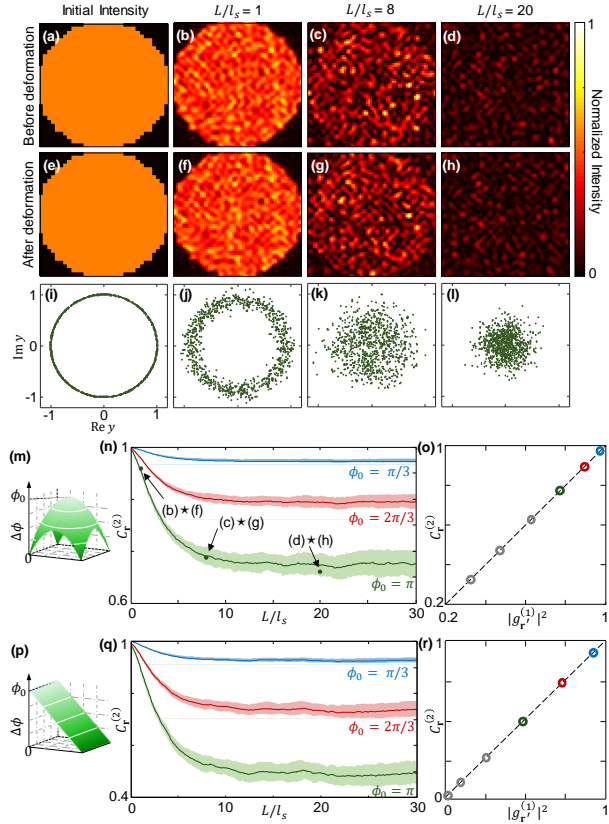

FIG 2. Simulation results with transmission matrix. Varying the optical thickness $L/l_s$, We calculated output intensity (a-d) before deformation and (e-h) after deformation where phase deformation has quadratic shape of (m) with maximum value $\phi_0 = \pi$. (i-l) Distribution

of the scattered optical fields $y$ before deformation (a-d). (n) Dependence of $C_\mathbf{r}^{(2)}$, correlation coefficient between speckles before and after deformation, on $L/l_s$. Three points represent 2-D correlation coefficients between corresponding images. (o) Dependence of saturated $C_\mathbf{r}^{(2)}$ in (n) on $|g_{\mathbf{r},\mathbf{r}'}^{(1)}|^2$, or squared modulus of field correlation between initial fields. Dotted line represents theoretical values from (9). (p) Phase deformation of tilted planar shape with maximum value $\phi_0$. With (p), the same procedures are repeated in (q) and (r).

Based on the simulated TMs, we obtain transmitted output speckle fields and their correlation coefficients, as shown in Fig. 2. Here an incident beam with uniform intensity and a circular boundary is assumed, which describes a beam passed through a circular aperture [Fig. 2(a)]. Figs. 2(b)–(d) are the output speckle intensity patterns when the initial beam transmits through complex media where $L/l_s$ is 1, 8, and 20. In a weakly scattering media [$L/l_s = 1$, Fig. 2(b)], the speckle intensity is negligible beyond the circular region because scattering is not sufficient to diffuse the input beam. As the optical thickness increases, the output beam diffuses more strongly, spreading over the entire area, and the overall intensity of speckle patterns is diminished according to Ohm's law. Then, we applied a phase deformation to the original incident beam, which is not directly revealed in the intensity profile before passing the turbid layer [Fig. 2(e)]. Here, the shape of deformation $\Delta\phi$ is followed a quadratic function as shown in Fig. 2(m) where the maximum phase difference $\phi_0$ is $\pi$ rad. Figs. 2(f)–(h) exemplify the apparent change in speckle patterns after the phase change of the incident beam, showing the effect of a scattering layer. Figs. (i)–(l) show the distribution of optical field $y$ before deformation corresponding to Figs. (a)–(d). We set an initial field to have uniform amplitude with random phase, which is depicted as a circle [Fig. 2(i)]. As shown in Figs 2(j)–(k), the variance of amplitude is increased as the scattering layer gets thickened. Eventually, the field distribution becomes fully random, which is described as complex Gaussian, in the presence of strong scattering [Fig. 2(l)].

For further investigations, we calculated the intensity correlation coefficient $g_\mathbf{r}^{(2)}$ to quantify the speckle change induced by the incident wavefront deformation in the numerical simulations. For convenience, we defined $C_\mathbf{r}^{(2)}$ or intensity correlation coefficient as $g_\mathbf{r}^{(2)} - 1$. Firstly, $C_\mathbf{r}^{(2)}$ between the three pairs of speckle images in Fig. 2 is 0.963, 0.787, and 0.760 when $L/l_s$ is 1, 8, and 20 respectively, and we depict these values as three points in Fig. 2(n). The cross signs in their tags mean a 2-D correlation between two specified images. We extended our approach to continuously varying $L/l_s$ from 0 to 30. Fig. 2(n) shows the change of $C_\mathbf{r}^{(2)}$ with three different phase deformations whose shape follows the quadratic function in Fig. 2(m) with corresponding $\phi_0$. We repeated 50 iterations for each curve, and the standard deviations are represented by the shaded area. In all the curves, $C_\mathbf{r}^{(2)}$ drops gradually starting from unity as the scattering layer gets thicker, but there is no further decline of $C_\mathbf{r}^{(2)}$ if the optical thickness exceeds a criterion of $L/l_s > 10$.

Figure 2(o) shows the comparison of the saturated value of $C_\mathbf{r}^{(2)}$ with $|g_{\mathbf{r},\mathbf{r}'}^{(1)}|^2$ for various quadratic wavefront deformations. While only three cases of phase deformation are shown in Fig. 2(n), we conducted further experiments for three other phase deformations $\phi_0 = 4\pi/3$, $5\pi/3$, and $2\pi$. These results are plotted with solid circles. The exact linear relationship between $C_\mathbf{r}^{(2)}$ with $|g_{\mathbf{r},\mathbf{r}'}^{(1)}|^2$ clearly shows the validity of the theoretical result in Eq. (6).

To verify that the universality is independent of the shape of the deformed wavefront, we repeated these procedures with a tilted wavefront as described in Fig. 2(p), and the results are shown in Fig. 2(q). Notably, we could observe the coincidence of $|g_{\mathbf{r},\mathbf{r}'}^{(1)}|^2$ and saturate $C_\mathbf{r}^{(2)}$ for the tilted wavefront as well [Fig. 2(r)].

The obtained result can be interpreted to mean that $C_\mathbf{r}^{(2)}$ converges to $|g_{\mathbf{r},\mathbf{r}'}^{(1)}|^2$, or the absolute square of the field correlation between the initial and deformed wavefronts before the scattering. This is because light transported in a complex media conveys the field correlation before scattering to the intensity correlation after scattering. This originates with the linear nature of a scattering matrix, which only transforms the basis of optical modes. The linearity also ensures that the correlation between speckles is limited by the pre-scattered field correlation, so $C_\mathbf{r}^{(2)}$ remains saturated despite a scattering medium with large optical thickness.

We also emphasize that the onset of the saturation of $C_\mathbf{r}^{(2)}$ occurs when the wavefront is scrambled to satisfy the condition of Reed's theorem. The saturation of $C_\mathbf{r}^{(2)}$ is reached when the optical thickness is increased enough so that the spatial distribution of the scattered field follows complex Gaussian as shown in Fig. 2(l). This condition coincides to when the intensity distribution follows Rayleigh statistics, which is the prominent characteristic of speckles.

## IV. EXPERIMENTAL SETUP

To verify experimentally, we conducted an experiment on speckle decorrelation induced by the morphological deformation of a microfluidic channel using a system in Ref. [24]. The principle of the experiment is illustrated in Fig. 3. When a coherent laser beam passes through a transparent microfluidic channel made of PDMS (Polydimethylsiloxane), the wavefront is deformed according to the refractive index difference and the geometry of the channel [Fig. 3(a)]. When positive pressure is applied inside the channel, its internal wall is inflated and the wavefront is deformed accordingly [Fig. 3(b)]. Then, a scattering layer below the channel visualizes the difference in wavefronts using the change in speckle, as illustrated in Fig. 1(a).

To implement the scheme, we composed the setup described in Fig. 3(c). The internal pressure of a microfluidic channel was controlled by changing the volume inside a syringe with the aid of a syringe pump (PHD ULTRA CP 4400, Harvard Apparatus, USA). The pressure inside the channel is monitored with a reference sensor connected to the body of the syringe. We used a channel which exhibited both transparency and elasticity, and the dimension of the channel was 4 mm (width) × 100 μm (height) × 30 mm (length), as depicted in the inset of Fig. 3(c). The two terminals of the channel are connected to the syringe using a T-shaped connector so that the pressure controlled by the syringe is directly and equally applied to both ends of a microfluidic channel.

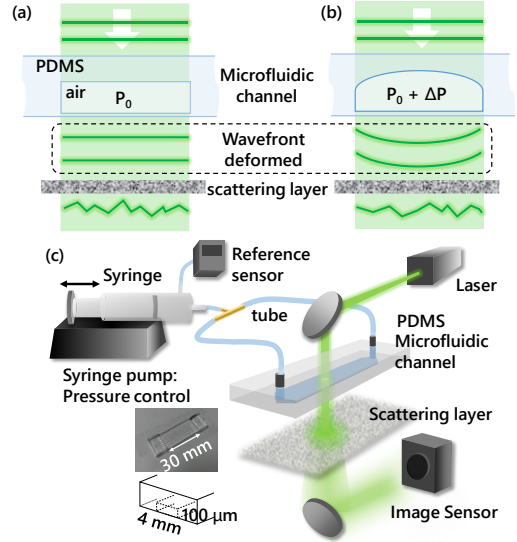

FIG. 3. Experimental scheme to deform the wavefront of a beam. (a) The laser beam passing through a microchannel and a scattering layer. (b) Change of wavefront and speckle pattern due to the inflated geometry of the channel with increased internal pressure. (c) Experimental setup. We controlled the pressure with a syringe pump.

In the optical setup, a diode-pumped solid state laser ($\lambda$ = 532 nm, Shanghai Dream Laser Co., Shanghai, China) was used as a coherent light source. The laser beam impinges onto the surface of the microfluidic channel. The beam which is transmitted through the channel encounters a scattering layer after a propagation of 2 cm. Speckle pattern formed by the scattering layer is recorded with a CCD image sensor (INFINITYlite, Lumenera, USA). The separation between the layer and the image sensor was adjusted to achieve the highest signal-to-noise ratio.

## V. EXPERIMENTAL RESULTS

Using the setup, we recorded output intensity images of speckle patterns and analyzed their spatial correlations when a positive pressure was applied to the microfluidic channel. First, the original beam and deformed beams were imaged without a scattering layer [Figs. 4(a)–(b)]. After the deformation, the beam intensity did not show significant changes, although the internal pressure $P$ inside the microfluidic channel was increased to 3 kPa. The correlation between the two images was 0.971. However, when a scattering layer diffused the beam and resulted in speckle formation, as shown in Figs. 4(c)–(h), it resulted in significant changes in intensity patterns.

To study the effects of scattering media, we used three types of scattering samples to vary the scattering strength: a single layer of a translucent tape (Scotch Magic Tape, 3M, United States) [Figs. 4(c)–(d)], ten layers of the Scotch tape [Figs. 4(e)–(f)], and a 15° diffuser (#54-495, Edmund Optics, United States) [Figs. 4(g)–(h)]. The picture of a hole on the optics table seen through each scattering sample is displayed at the top of the figures. Comparing the speckles before and after wavefront deformation, a discernable change in speckle pattern is found for every sample. Also, the correlation between those images was significantly reduced to 0.393, 0.387, and 0.392 for one layer, ten layers of Scotch tape, and 15° diffuser, respectively. The similar values imply that $C_r^{(2)}$ is independent of the thickness or property of a scattering layer.

In addition, we applied a continuous increase of pressure and calculated the variation of output intensity correlation [Fig. 4(i)]. The pressure was increased from air pressure until it reached the maximum limit of the reference sensor. In the figure, the error bars represent the standard deviation from 5 repeated measurements. Also, $C_r^{(2)}$ calculated from Figs. 4(a)–(h) are indicated by the arrows.

The curves in Fig. 4(i) show a decline of $C_r^{(2)}$ due to the increased deformation with increasing $P$. $C$ drops much faster with a scattering layer than without it. When $P$ is 3.2 kPa, $C_r^{(2)}$ decreased from unity to 0.97 without a scattering layer, while it was decreased to approximately 0.36 with the layer, showing a 21-fold enhancement in sensitivity. Importantly, the decorrelation graphs coincide regardless of the thickness or type of the scattering layer. The p-value of $C_r^{(2)}$ between one layer and ten layers of Scotch tape was 0.4, and between one layer of Scotch tape and 15° diffuser it was 0.5, which proves there is no significant statistical difference among them.

We note that in our experiment the scattered field can be affected by other factors such as a distance between the layer and the image sensor due to diffraction. However, since the measured intensity probability distribution follows Rayleigh statistics, it is inferred that the field probability distribution follows complex Gaussian, which allows the application of Reed's theorem.

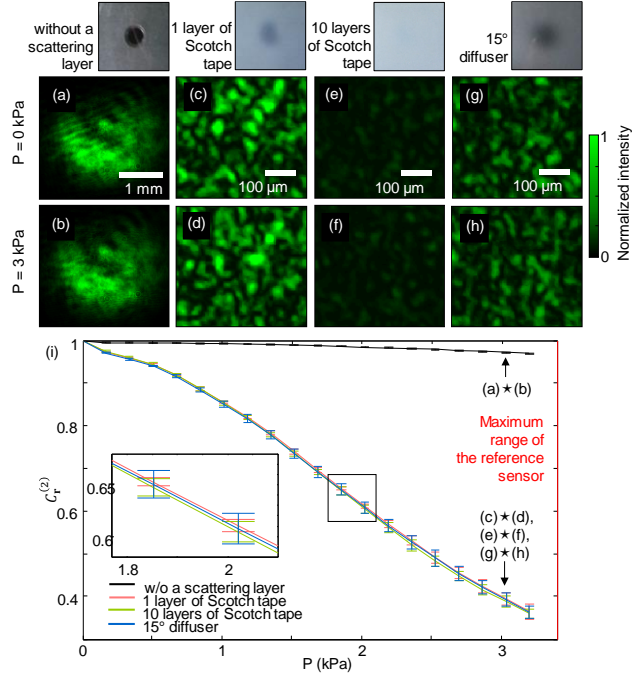

FIG. 4. Intensity images with various scattering media: (a–b) none, (c–d) 1 layer, (e–f) 10 layers of Scotch tape, and (g–h) 15° diffuser, when the internal pressure of the microfluidic channel was at air pressure or 3 kPa. The images above them show the same object, a hole in an optics table, seen through each scattering sample. (i) Change in correlation coefficient $C_r^{(2)}$ when the pressure is increased. The inset represents magnification of the rectangular box in the original figure.

## VI. DISCUSSION AND CONCLUSIONS

In this letter, we present the universal sensitivity of speckle intensity correlations to wavefront deformation in light diffusers. When the alteration in speckle intensities is expressed as the correlation coefficient, it was analytically shown that the speckle change is relevant to just the wavefront deformation whenever a scattering layer belongs to the diffusive regime. Employing a numerical simulation using a TM, we found that the consistency between the correlation and the saturated value exactly matches the modulus square of the incident field correlation. With a wavefront-deforming setup using a deformable microchannel, we experimentally confirmed the universality of the spatial intensity correlation to the changes in the wavefront of a beam passing through various scattering layers, regardless of the type and thickness of the scattering layer.

The present work may open new avenues for optical metrology and interferometry because it directly enables a simple speckle-based approach for the detection of wavefront change using a scattering layer. Importantly, the present work provides a theoretical framework and experimental verification for general speckle metrology. In particular, the present work provides a means to retrieve phase images directly regardless of scattering media and without any calibration or characterization of the scattering media in use. From a practical point of view, our concept is easily implemented with the insertion of a scattering medium in the midst of a beam trajectory, without the need for a reference beam, to obtain the interference pattern between speckles. It also provides a method of cost-effective remote sensing, because the setup only requires an ordinary scattering layer, like the translucent Scotch tape we used. Our principle is applicable independent of the choice of scattering medium, guaranteeing that any diffusive layer always gives the same expected performance at given conditions. We expect this approach will find direct applications in various applications where the optical phase change is of interest, ranging from metrology, analytical chemistry, sensing, and biomedical optics.

## ACKNOWLEDGEMENTS

This work was supported by KAIST, and the National Research Foundation of Korea (2015R1A3A2066550, 2014K1A3A1A09063027, 2012-M3C1A1-048860, 2014M3C1A3052567) and Innopolis foundation (A2015DD126).